\documentclass[aps,pra,twocolumn,superscriptaddress,floatfix]{revtex4}

\usepackage{graphicx}  
\usepackage{amsmath}   

\begin{document}

\widetext

\title{Search for a liquid-liquid critical point in models of silica}

\author{Erik Lascaris}
\affiliation{Center for Polymer Studies and Department of Physics,
  Boston University, Boston, MA 02215 USA} 

\author{Mahin Hemmati}
\affiliation{Department of Chemistry and Biochemistry, Arizona State
  University, Tempe, AZ 85287 USA} 

\author{Sergey V. Buldyrev}
\affiliation{Department of Physics, Yeshiva University, 500 West 185th
  Street, New York, NY 10033 USA} 

\author{H. Eugene Stanley}
\affiliation{Center for Polymer Studies and Department of Physics,
  Boston University, Boston, MA 02215 USA}

\author{C. Austen Angell}
\affiliation{Department of Chemistry and Biochemistry, Arizona State
  University, Tempe, AZ 85287 USA}

\date{5 May 2014}

\begin{abstract}

\noindent
Previous research has indicated the possible existence of a liquid-liquid
critical point (LLCP) in models of silica at high pressure.  To clarify this
interesting question we run extended molecular dynamics simulations of two
different silica models (WAC and BKS) and perform a detailed analysis of the
liquid at temperatures much lower than those previously simulated.  We find no
LLCP in either model within the accessible temperature range, although it is
closely approached in the case of the WAC potential near 4000~K and 5~GPa.
Comparing our results with those obtained for other tetrahedral liquids, and
relating the average Si-O-Si bond angle and liquid density at the model glass
temperature to those of the ice-like $\beta$-cristobalite structure, we
conclude that the absence of a critical point can be attributed to insufficient
``stiffness'' in the bond angle.  We hypothesize that a modification of the
potential to mildly favor larger average bond angles will generate a
LLCP in a temperature range that is accessible to simulation.  The tendency to
crystallize in these models is extremely weak in the pressure range studied,
although this tendency will undoubtedly increase with increasing stiffness.

\end{abstract}

\maketitle

\section{Introduction}

\noindent
Silica (SiO$_2$) is one of the most important and widely used materials in
today's world.  One could say that the fact of its ubiquity is as clear as
window glass.  Because silica is an excellent insulator and can be easily
created through thermal oxidation of the silicon substrate, SiO$_2$ is also the
insulator of choice in the semiconductor industry.  Optical fibers made from
pure silica are widely used by the telecommunications industry and, because
silica and silicates make up over 90\% of the Earth's crust, SiO$_2$ plays a
major role in the geosciences.

Liquid silica is the extreme case of a ``strong'' liquid. When cooled, its
viscosity approaches the glass transition slowly, following the Arrhenius law
$\log \eta \propto 1/T$.  In contrast, the so-called ``fragile'' liquids reach
this glass transition far more quickly.  Glasses rich in silica, but modified
by other oxides to lower their viscosities, are ``strong'' liquids that have
slow vitrification so are preferred by glassblowers who need time to work their
magic.

Simulations have indicated that liquid silica does not behave like a strong
liquid for all temperatures, however.  Using the BKS model
\cite{vanBeestPRL1990} (see Appendix~A), Vollmayr {\it et al.} found that at
very high temperatures the diffusion greatly deviates from the Arrhenius law
(and thus behaves like a fragile liquid), and that the temperature-dependence
of the diffusion better fits the Vogel-Fulcher law \cite{VollmayrPRB1996}.  It
was later shown by Horbach and Kob \cite{HorbachPRB1999} that the
temperature-dependence can also be fitted well by a power law of the shape $D
\propto (T - T_{\text{MCT}})^{\gamma}$ in which the exponent $\gamma$ is close
to 2.1 (compared to 1.4 for water) and $T_{\text{MCT}} \approx 3330$~K.  This
temperature dependence is often found in simple liquids and has been described
in terms of mode-coupling theory (MCT) \cite{GotzeLHSS1989,GotzeRPP1992}.  A
deviation from the Arrhenius law has also been measured in other models of
silica \cite{HemmatiMineralogy2000}, and small deviations from a pure Arrhenius
law were found for the viscosity in experimental data
\cite{HessCG1996,RosslerJNCS1998}.  This transition from fragile to strong upon
cooling (often called the ``fragile-to-strong crossover'') has also been found
in simulations of other tetrahedral liquids, such as BeF$_2$
\cite{AngellPCCP2000}, silicon \cite{SastryNatM2003,AshwinPRL2004}, and water
\cite{ItoNat1999, GalloJCP2012, XuPNAS2005}.  This phenomenon is not restricted
to tetrahedral liquids, however.  For example, it has been proposed that the
fragile-to-strong crossover might be a behavior common to all metallic
glass-forming liquids \cite{ZhangJCP2010,LadJCP2012}.

In addition to the fragile-to-strong crossover, it has been proposed that
liquid silica also has a liquid-liquid critical point (LLCP)
\cite{PoolePRL1997, SaikaVoivodPRE2000, AngellAIP2013} much like that proposed
for liquid water \cite{PooleNat1992}.  These phenomena may be related.  It was
recently shown that in analog plastic crystal systems many strong glass-formers
are accompanied by a singularity (a lambda-type order-disorder transition) at
high temperatures, and that in silica this singularity could be a LLCP
\cite{AngellAIP2013}.  The fragile-to-strong crossover arises simultaneously
with a large increase of the isobaric heat capacity $C_P$.  If a LLCP exists in
silica, this heat capacity maximum should have its origin in its critical
fluctuations.  The discovery of a LLCP in liquid silica would thus provide a
unifying thermodynamic explanation for the behavior of liquid silica.

\section{Methods}
\label{SEC:Methods}

\noindent
We consider here two different models of silica, the BKS model by van Beest
{\it et al.} \cite{vanBeestPRL1990} and the WAC model (also known as the TRIM
model for silica) introduced by Woodcock {\it et al.}  \cite{WoodcockJCP1976}.
Both models represent SiO$_2$ as a simple 1:2 mixture of Si ions and O ions,
i.e., without any explicit bonds.  One difference between the two models is
that WAC uses full formal charges while in BKS partial charges are used.  For a
detailed description of both models, see Appendix A.

All simulations are done using Gromacs~4.6.1 \cite{Gromacs4}, with $N=1500$ ions,
using the Ewald sum (PME) for electrostatics, and the v-rescale thermostat
\cite{BussiJCP2007} to keep the temperature constant.  Most simulations are
done in the constant-volume/constant-temperature ($NVT$) ensemble.  For the few
constant-pressure ($N\!PT$) simulations we use the Parrinello-Rahman barostat.
For most of the simulations we use a time step of 1~fs, but at very low
temperatures we increase the time step to 4~fs to speed up the simulations to
approximately 250~ns/day.  We carefully check the temperatures below which the
4~fs time step gives the same results as the 1~fs time step and do not include
any 4~fs data that lead to a small difference in pressure, energy, or
diffusion.

As a measure of the equilibration time, we define $\tau$ as the time at which
$\sqrt{\left< r_{\text{O}}(t)^{2} \right>} = 0.56$~nm, i.e., the average time
it requires for an O ion to move twice its diameter of 0.28~nm.  Most
simulations run for over $10\,\tau$, well beyond the time necessary for the
system to reach equilibrium.  For the range of temperatures and pressures
considered here, the root mean squared displacement of the O ion is roughly
1.1--1.6 times that of the Si ion, this factor being the largest at low
temperatures and low pressures.

An important structural feature is the coordination number of Si by O, since a
tetrahedral network is defined by 4-coordination of the network centers.  We
calculate the Si coordination number by the usual method, integrating the Si-O
radial distribution function up to the first minimum.  For both models, and at
all state points considered here (below 10~GPa), the coordination number lies
between 4.0 and 4.9.  The coordination number is the largest at high densities,
and levels off to 4 when the density is decreased and the pressure reaches zero
and becomes negative.

\section{Isochores}

\begin{figure}[!t]
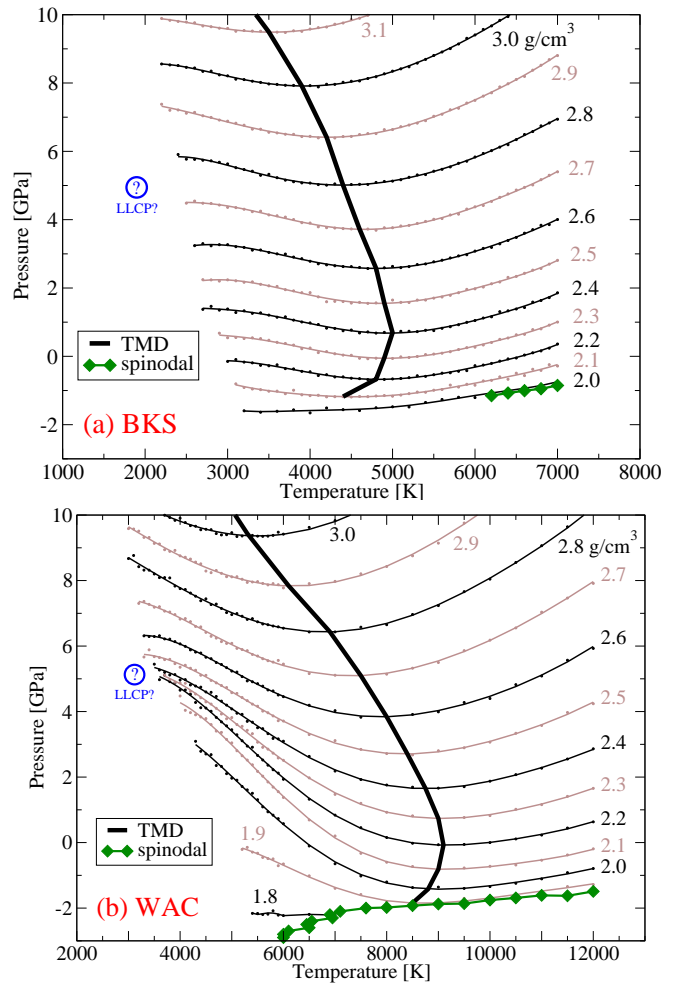
 
\centering
\includegraphics[angle=0,width=\linewidth]{isochores-BKS.eps} \\
\includegraphics[angle=0,width=0.95\linewidth]{isochores-WAC.eps}
\caption{ Isochores of liquid BKS silica (panel a) and liquid WAC
  silica (panel b).  Thin black/brown lines are the isochores, the
  temperature of maximum density (TMD) is indicated by a thick black
  line, and green diamonds indicate part of the liquid-vacuum
  spinodal.  Blue question marks indicate the approximate locations where a
  LLCP has been predicted by previous studies
  \cite{SaikaVoivodPRE2000, AngellAIP2013}.  The location of a LLCP
  can be identified by where the isochores cross.  It seems a LLCP in
  BKS is unlikely, as the isochores do not approach each other.  The
  isochores in WAC do approach each other, and might converge at the
  predicted point.  However, at low temperatures the isochores near
  2.3~g/cm$^3$ obtain a negative curvature.  If this curvature becomes
  more negative as $T$ goes down, then it is possible that the
  isochores will not cross below 3500~K.  We conclude that for the
  temperatures currently accessible, the isochores alone are
  insufficient to demonstrate a LLCP in WAC.  }
\label{FIG:isochores_BKS_WAC}
\end{figure}

\noindent
The most direct method of locating a critical point is to calculate the
pressure $P$ as a function of temperature $T$ along different isochores.  In a
$PT$-diagram the isochores cross within the coexistence region and at the
critical point.  At those state points (at a given $P$ and $T$) the system is a
combination of two different phases with different densities.  One can also
locate a critical point by plotting the isotherms in a $PV$-diagram in order to
determine the region in which the slope of the isotherms becomes zero (critical
point) or negative (coexistence region). Because it is easier to determine
whether two lines are crossing than whether a curve is flat, we study the
isochores.  Figure~\ref{FIG:isochores_BKS_WAC} shows the $PT$-diagrams with the
isochores of BKS and WAC.

Both diagrams are similar.  There is a clear density anomaly to the
left of the temperature of maximum density (TMD), and if we raise the
temperature by approximately 4000~K then the BKS isochores match those
of WAC reasonably well.  Thus, based on the isochores in
Fig.~\ref{FIG:isochores_BKS_WAC}, one could say that BKS and WAC are
very similar systems, and that they mainly differ in a shift of
temperature.

At very low $P$ and high $T$ the liquid phase is bound by the
liquid-gas (or liquid-vacuum) spinodal, and lowering $P$ below the
spinodal leads to spontaneous bubble formation.  At very low $T$ the
liquid becomes a glass, and the diffusion coefficient drops rapidly.
Because the time it takes to equilibrate the system is inversely
proportional to the rate of diffusion, simulations require too much
time once the oxygen diffusion $D_{\text{O}}$ drops below $\sim
10^{-8}$~cm$^2$/s, which is where the isochores stop in
Fig.~\ref{FIG:isochores_BKS_WAC}.  For both models this limit is
reached at a higher temperature for low $P$ than for high $P$.  This
is caused by the diffusion anomaly (an increase in $P$ leads to an
{\em increase\/} in diffusion), which is present in both BKS and WAC models.

No crystallization was observed, unless the pressure was raised to values far
outside the range of our detailed studies (e.g., above 40~GPa the WAC liquid
spontaneously crystallizes into an 8-coordinated crystal).  Normally,
crystallization is readily detected by a rapid drift of the energy to lower
values.  However, when the diffusivity is very low (as in the present system,
in the domain of greatest interest) the situation is different and crystal
growth can be unobservably slow.  More direct tests are then needed.  In the
present case we have sought information on crystal growth and melting by
creating a crystal front (half simulation box of the liquid interfacing with
half box of the topologically closest crystal) and have watched the crystal
front receding at high temperature.  However, the attempt to determine melting
point by lowering the temperature and observing reversal of the interface
motion, was unsuccessful because the growth rate became unobservably small
(observed over microseconds) before any reversal was seen.  We conclude that,
since this crystal front was put in by hand, the possibility of crystallization
by {\em spontaneous} nucleation (always the slowest step) followed by crystal
growth, is zero.

Based on the fitting and extrapolation of data, previous studies have
predicted a liquid-liquid critical point (LLCP) in both WAC and BKS
\cite{SaikaVoivodPRE2000}.  With the increase in computing power, and
using the techniques to speed up the simulations discussed in
Sec.~\ref{SEC:Methods}, we are able to obtain data at lower
temperatures than was previously possible.  Our results for BKS
(Fig.~\ref{FIG:isochores_BKS_WAC}a) show that for $T>2500$~K the
isochores are nearly parallel, and therefore a LLCP in BKS is very
unlikely.  On the other hand, the isochores of the WAC model
(Fig.~\ref{FIG:isochores_BKS_WAC}b) show a more interesting behavior
in that they clearly approach one another at low $T$ in the vicinity
of $P \approx 5$~GPa.

If we only consider the WAC isochores above 4000~K, then extrapolation
would predict that the isochores cross around 3500~K and 5~GPa.
However, below 4000~K we see that the isochores are starting to
display a negative curvature in the $PT$-plane.  This signals an
approach to a density minimum, which is the low-$T$ boundary of the
density anomaly region.  The negative curvature makes it hard to
perform an extrapolation that convincingly shows that the isochores
cross at lower $T$.  We can therefore only conclude that (for the
temperatures currently accessible) the isochores are insufficient to
prove or disprove the existence of a LLCP in WAC.

\section{Response functions}

\noindent
Upon approaching a critical point, the response functions should
diverge.  Although true divergence occurs only in the thermodynamic
limit $N \to \infty$, a large maximum should still be visible in response
functions such as the isothermal compressibility $K_T$ and the
isobaric heat capacity $C_P$ even when the box size is relatively
small.  Calculations using the Ising model and finite size scaling
techniques applied to simulation results have shown that (for
sufficiently large boxes) the location of the critical point is very
close to where both $K_T$ and $C_P$ reach their global maximum
\cite{KesselringJCP2013, LascarisAIP2013}.  If a LLCP truly exists in
WAC, then the $PT$-diagrams of $C_P$ and $K_T$ should show a large
$C_P$ maximum close to where $K_T$ has a maximum---exactly where the
isochores come together and where the LLCP has been predicted to be.

\begin{figure*}[!ht] 
\centering
\includegraphics[width=0.45\linewidth]{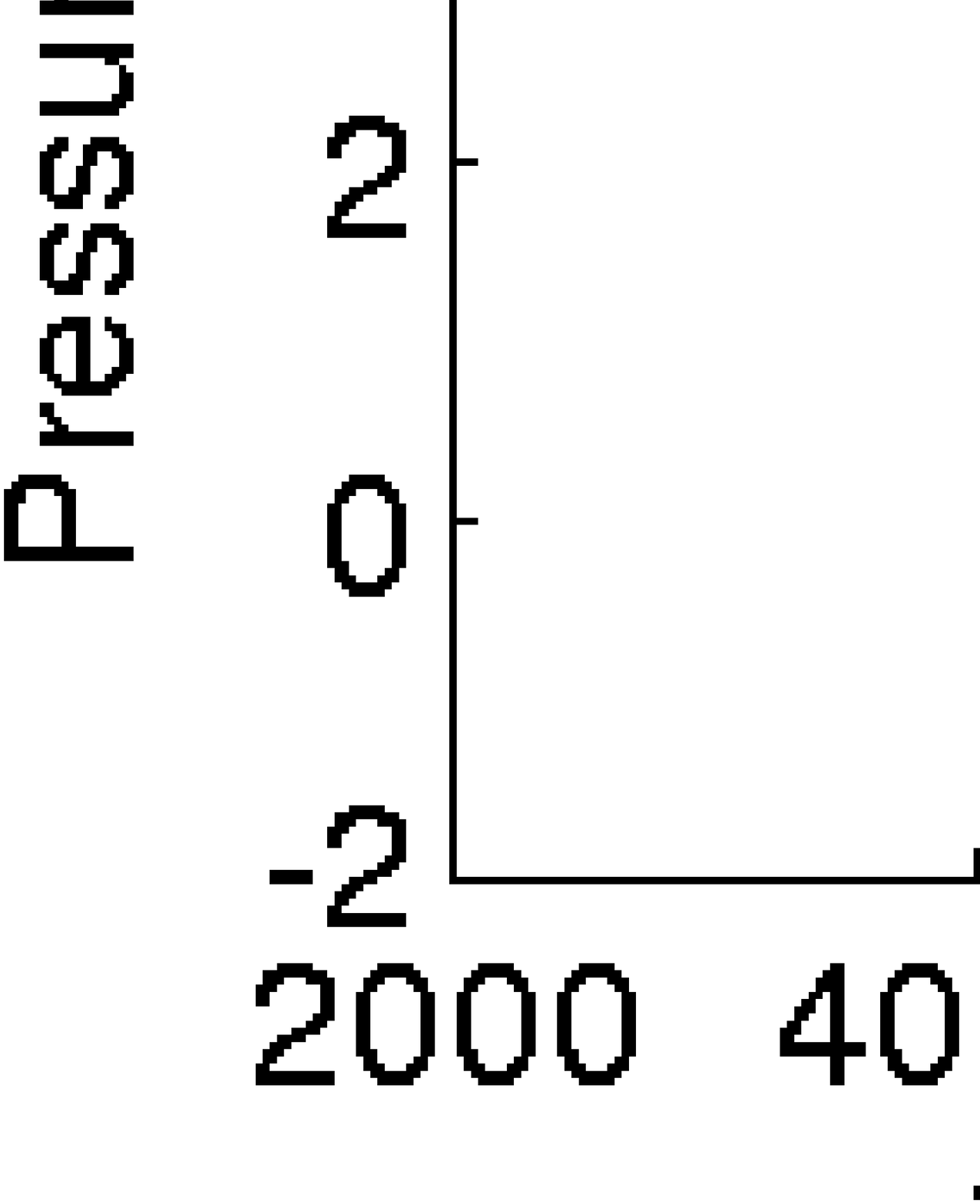}(a)
\hfill
\includegraphics[width=0.45\linewidth]{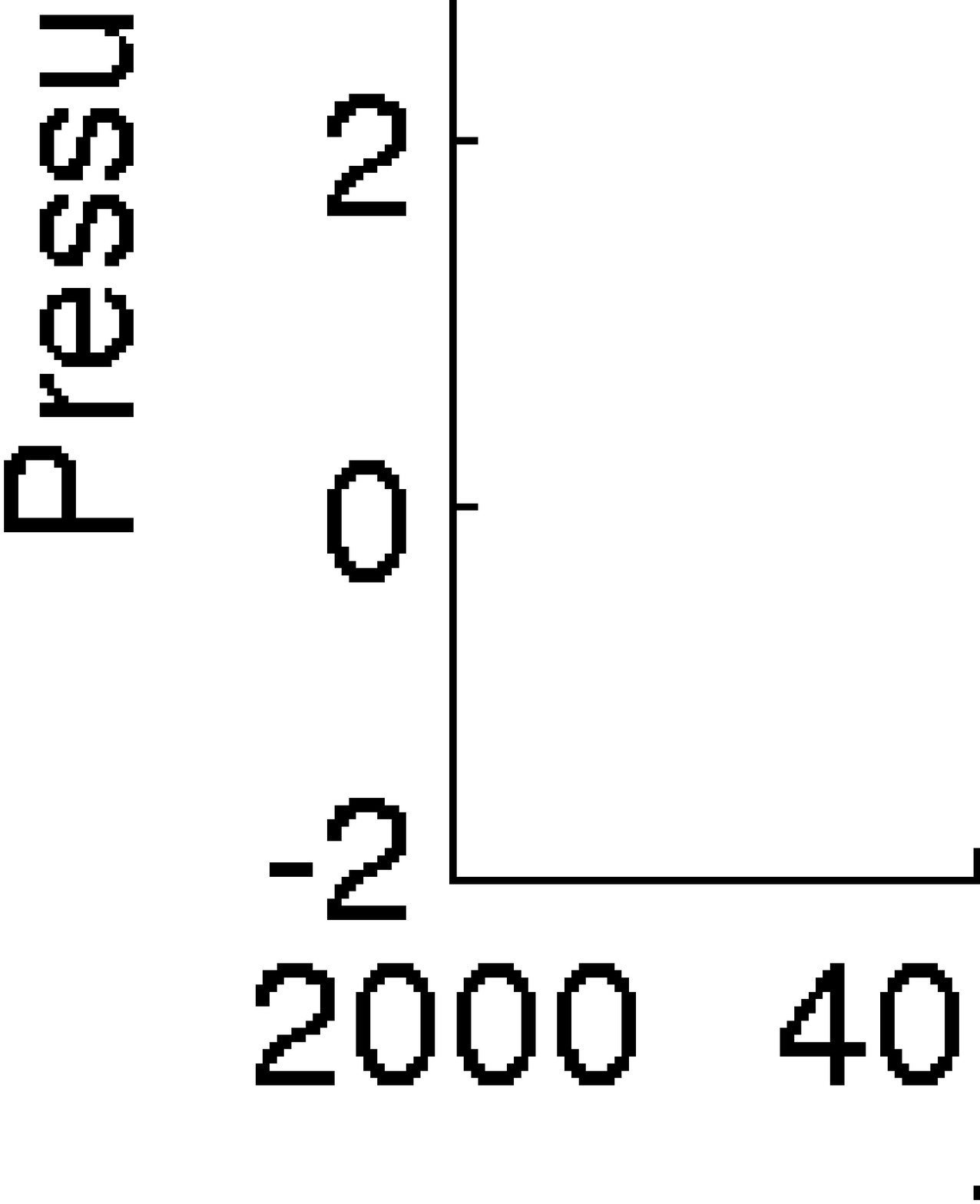}(b)
\quad \\
\includegraphics[width=0.45\linewidth]{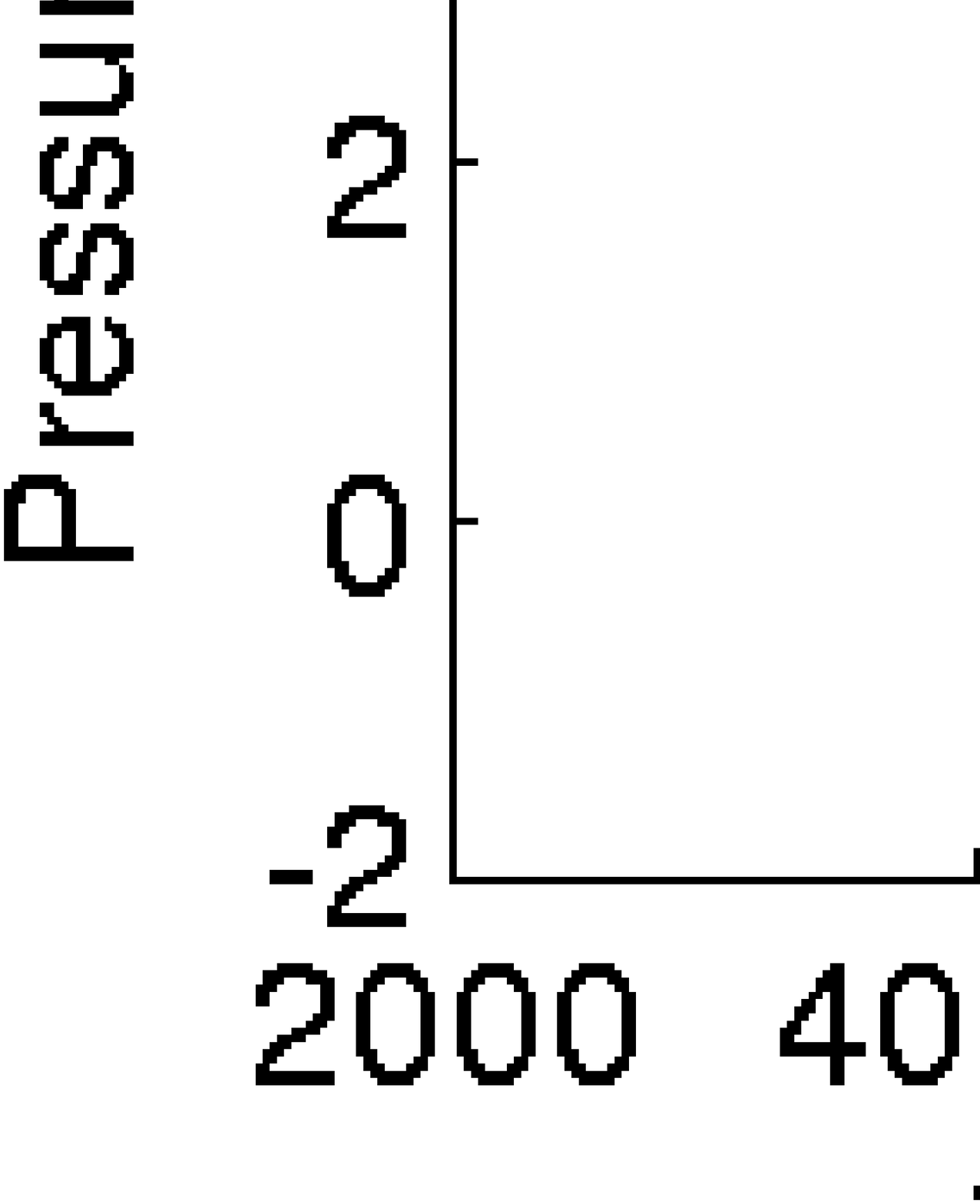}(c)
\hfill
\includegraphics[width=0.45\linewidth]{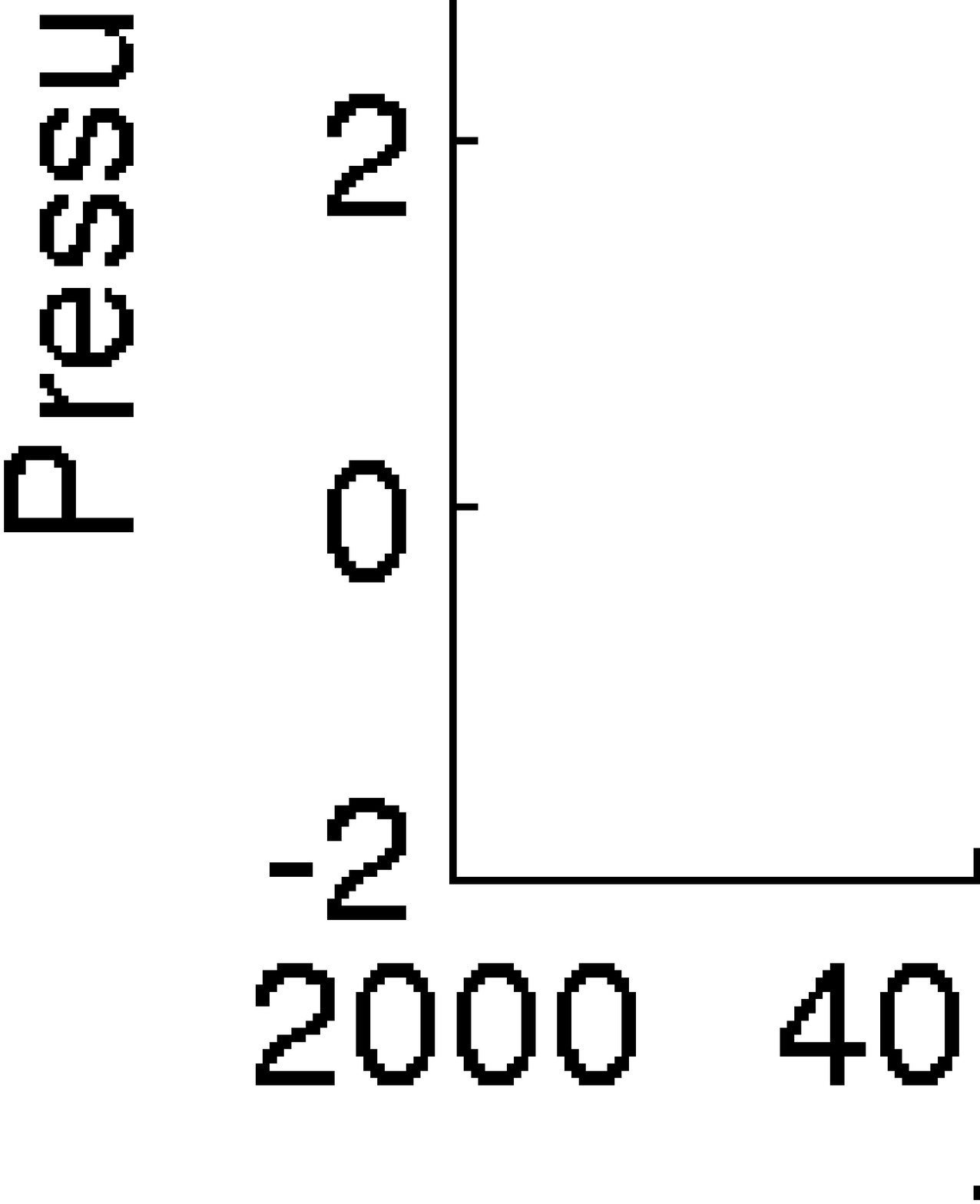}(d)
\caption{ Response functions of WAC. (a) The isothermal
  compressibility $K_T$ is consistent with a LLCP near 5~GPa, 4000~K
  because near that point $K_T$ has a global maximum.  (b) The
  isobaric heat capacity $C_P$, however, has a global maximum around
  1~GPa and 6000~K, far away from where $K_T$ has its global maximum.
  This is inconsistent with the LLCP hypothesis.  (c) The isobaric
  thermal expansivity $\alpha_P$ has its global minimum in between the
  global maxima of $K_T$ and $C_P$.  The contour line where
  $\alpha_P=0$ corresponds to the location of the TMD.  (d) The
  isochoric heat capacity $C_V$ has its global maximum the furthest
  away from the global $K_T$ maximum.  }
\label{FIG:response_functions_WAC}
\end{figure*}

Figure~\ref{FIG:response_functions_WAC} shows four response functions
for WAC: (a) the isothermal compressibility $K_T$, (b) the isobaric
heat capacity $C_P$, (c) the isobaric thermal expansivity $\alpha_P$,
and (d) the isochoric heat capacity $C_V$.  These have been obtained
using $NVT$ simulations together with the smooth surface technique
described in Appendix~\ref{appendix_smooth_surface}.  To check the
results generated by this technique, we determine whether the response
functions satisfy the thermodynamic relation $VT \alpha_{P}^{2}/K_T +
C_V - C_P = 0$.  Because of statistical errors in the data we find
slight deviations from zero, but these are less than 1~J/(mol~K) in
magnitude.

The compressibility $K_T$ in Fig.~\ref{FIG:response_functions_WAC}a shows a
clear global maximum near $P \approx 5$~GPa and $T \approx 4000$~K, because
this is where the isochores in Fig.~\ref{FIG:isochores_BKS_WAC}b are the
closest together in terms of pressure.  It is quite likely that below 4000~K
this maximum increases further.  If WAC has a LLCP then $C_P$ should also have
a maximum in that vicinity.  However, Fig.~\ref{FIG:response_functions_WAC}b
shows that this is not the case.  There is clear global $C_P$ maximum, but it
is located near $P \approx 1$~GPa and $T \approx 6000$~K, which is far from the
global $K_T$ maximum.  Therefore, based on the response functions, we conclude
that WAC does not have a LLCP.

The isobaric thermal expansivity $\alpha_P$
(Fig.~\ref{FIG:response_functions_WAC}c) has a global minimum between the
global maxima of $C_P$ and $K_T$ (Figs.~\ref{FIG:response_functions_WAC}a,b).
This should come as no surprise, since $C_P \propto \left< (\Delta S)^2
\right>$ arises from fluctuations in entropy and $K_T \propto \left< (\Delta
V)^2 \right>$ from volume fluctuations, while the expansivity $\alpha_P \propto
\left< \Delta S \Delta V \right>$ arises from a combination of both.  Even
though the global maxima occur at different places, the slopes $dP/dT$ of the
loci of local maxima are the same, so it seems likely they have a common
origin.  Because the system is not quite critical, the enthalpy fluctuations
that determine the heat capacity can be statistically independent of the
density fluctuations.

\begin{figure}[!ht]
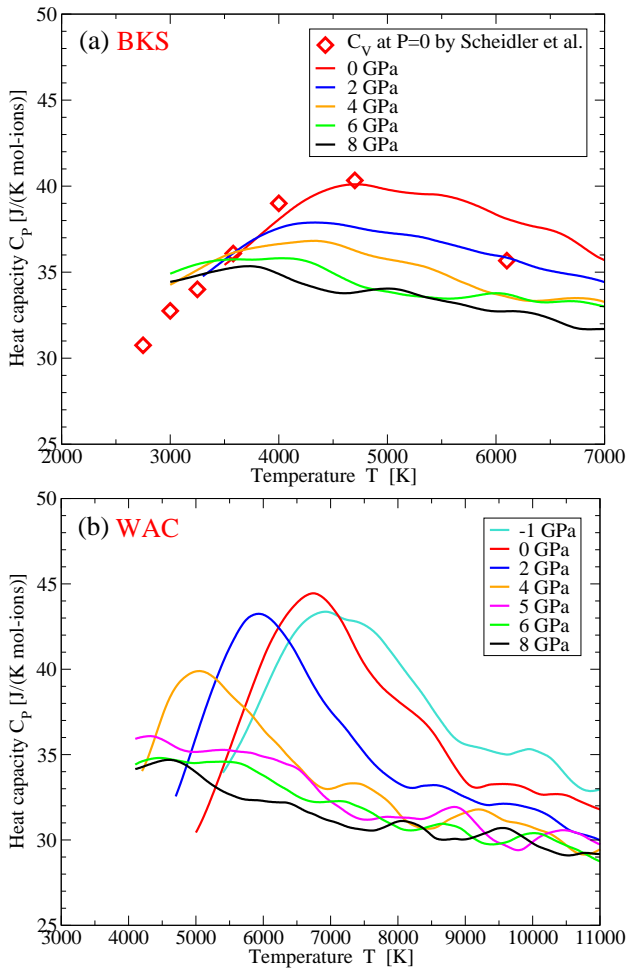
 
\centering
\includegraphics[angle=0,width=0.95\linewidth]{Cp-vs-T_smoothingspline_BKS.eps}
\includegraphics[angle=0,width=0.95\linewidth]{Cp-vs-T_smoothingspline_WAC.eps}
\caption{ Comparison of the heat capacities of BKS (panel a) and WAC (panel b),
  obtained by calculating the smoothing spline of $H(T)$ at constant $P$,
  followed by taking its derivative (a slightly different method than was
  used in Fig.~\ref{FIG:response_functions_WAC}b).  At 8~GPa there is no
  significant difference between the WAC and BKS models, but below 5~GPa WAC
  has a large maximum in the range 5000--8000~K (also clearly visible in
  Fig.~\ref{FIG:response_functions_WAC}b).  In panel b we have included $C_V$
  data of Scheidler {\it et al.} \cite{ScheidlerPRB2001} (red diamonds),
  which shows a maximum around 4500~K.  Near the TMD (around 5000~K for
  $P=0$) the expansivity is small, which means that $C_V \approx C_P$, in
  agreement with our results.  For BKS this maximum is less clear in $C_P$,
  though still visible.  Because of small fluctuations in the data, it is
  difficult to obtain a fit of $H(T)$ that produces a perfect estimate of
  $C_P = dH/dT$, leading to artificial oscillations in $C_P$.  A larger data
  set would reduce this artifact.  In addition, the smoothing spline method
  assumes zero curvature at the end-points of the data, and this leads to
  artifacts at very low $T$ and very high $T$.  For clarity, we have removed
  the parts of the curves below the temperature at which $C_P$ starts to bend
  toward a constant a $C_P$ value.  }
\label{FIG:smoothingspline_BKS_WAC}
\end{figure}

The variation of the heat capacity with temperature at constant pressure is
shown over the temperature range in which the system remains in equilibrium, in
Fig.~\ref{FIG:smoothingspline_BKS_WAC}.
Fig.~\ref{FIG:smoothingspline_BKS_WAC}b is basically a cross-section of
Fig.~\ref{FIG:response_functions_WAC}b.  We note first that at moderately high
pressures, 8~GPa, there is no difference between the WAC and BKS models.  In
each case the heat capacity reaches about 35~J/(K~mol) before the diffusion
becomes too slow that we can no longer equilibrate.  This is 1.4 times the
vibrational heat capacity of $3R \approx 25$~J/(K~mol), as is typical of
moderately fragile inorganic liquids (e.g.  anorthite, ZnCl$_2$) right before
ergodicity is broken \cite{AngellJNCS1985,HemmatiJCP2001}.  However, at
pressures between zero and 5~GPa, a major difference is seen between the
models.

Near the TMD we have $C_P \approx C_V$ (because the expansivity is very small)
so we can compare data with $C_V$ from Scheidler {\it et al.}
\cite{ScheidlerPRB2001} for the case of BKS at $P=0$. The agreement is
quantitative, up to the point where the earlier study was cut off.  Our data
confirms the existence of a peak in the equilibrium heat capacity---an unusual
behavior that was not reported in Ref.~\cite{ScheidlerPRB2001} but had been
noted in the earlier study of Saika-Voivod {\it et al.}
\cite{SaikaVoivodNat2001} and was emphasized in Ref.~\cite{AngellAIP2013}.

Although BKS is far from having a critical point, the existence of this $C_V$
maximum reveals the tendency of this system---which accords well with many
aspects of experimental silica---to develop the same anomalous entropy
fluctuations, and an analog of the Widom line made famous by water models.

For the WAC model (which approaches criticality much more closely than BKS
does, as we have already seen in Fig.~\ref{FIG:isochores_BKS_WAC}), this heat
capacity peak becomes much more prominent, reminiscent of the behavior of the
Jagla model near its critical point.  $C_P$ reaches a value almost twice that
of the vibrational component; behavior unseen in any previous inorganic system
except for BeF$_2$ which is a WAC silica analog \cite{HemmatiJCP2001}.

\section{Discussion}

\noindent
We find no LLCP in either model within the accessible temperature range,
although it is closely approached in the case of the WAC potential near 4000~K
and 5~GPa.  The isochores of BKS, which are the most direct indicators of
criticality in a physical system, fail to converge into a critical point.  In
the case of WAC we cannot conclude anything from the isochores, but an analysis
of the global extrema of the response functions indicates that there is no LLCP
in WAC because the global $C_P$ maximum and the global $K_T$ maximum are
significantly separated in the $PT$-plane.

Liquid silica forms a tetrahedral network of bonds, and below we will show that
the lack of a LLCP is related to the openness of this network structure, which
in turn is related to the stiffness of the inter-tetrahedral bond angles.  In
addition we will argue that criticality in WAC could be achieved with an
adaptation of the pair potential.

The occurrence of a LLCP requires two competing liquid structures that can be
in a (meta-stable) equilibrium with each other.  In the case of a tetrahedral
network-forming liquid the two relevant structures are usually: (i) a
high-density collapsed structure that is highly diffusive, and (ii) a
low-density open network structure that is more rigid, i.e., one that is still
a liquid but less diffusive and more structured.  Because the high-density
structure occupies a smaller volume but has higher entropy (more disorder), the
competition between these two structures is accompanied by a region with a
density anomaly: $\alpha_P \propto \left<\Delta S \Delta V\right> <0$.

The high-density structure is very stable and is the dominant structure at high
temperatures, but the low-density structure requires a more delicate balance of
forces in order to be stable.  If the bonds in the liquid are too flexible, the
liquid collapses into the high-density structure.  On the other hand, if the
bonds are too rigid the liquid can no longer flow and
becomes a glass.

\begin{figure}[!htbp]
\centering
\includegraphics[angle=0,width=0.8\linewidth]{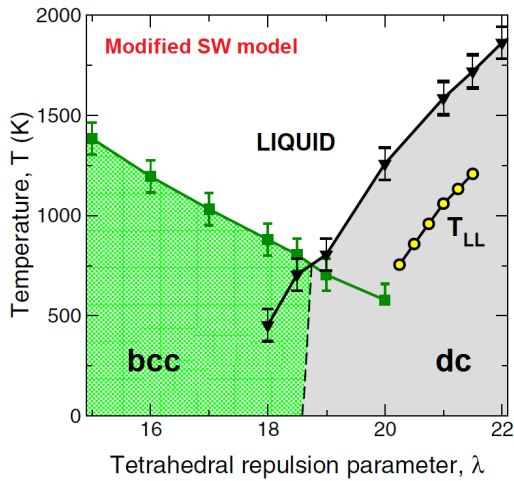}
\caption{ Phase diagram of the modified Stillinger-Weber potential in
  terms of the tetrahedral repulsion parameter $\lambda$ and
  temperature $T$, at zero pressure
  \cite{MolineroPRL2006}.
  The black triangles indicate the melting
  line of the diamond cubic (dc) crystal, while the green squares
  denote the melting line of the bcc crystal.  The dashed line
  separates the dc and bcc regions.  Yellow circles indicate the
  transition temperature $T_{\text{LL}}$ at which the liquid-liquid
  phase transition line crosses the $P=0$ isobar for that particular
  value of $\lambda$.  Silicon is represented by $\lambda=21$ and has
  a liquid-liquid critical point at $-0.60$~GPa \cite{VasishtNP2011},
  and therefore all LLCPs for $\lambda>20.25$ lie at negative
  pressures (there is a LLCP for each value of $\lambda$).  For
  $\lambda<20.25$ the LLCPs are at positive pressures and therefore
  the phase transition line can no longer be seen in this diagram.
  When $\lambda$ is large the system easily crystallizes, and
  therefore the phase transition line at $P=0$ can no longer be
  accurately located when $\lambda > 21.5$.  }
\label{FIG:modSW}
\end{figure}

There are several studies that address this situation.  The 2006 study of
Molinero {\it et al.} \cite{MolineroPRL2006} shows how reducing the three-body
repulsion parameter $\lambda$ in the Stillinger-Weber potential
\cite{StillingerPRB1985} (which controls the bond angle stiffness) causes the
first order liquid-liquid phase transition of silicon ($\lambda = 21$) to
disappear at $P=0$ when $\lambda < 20.25$ (see Fig.~\ref{FIG:modSW}).  This
transition occurs between a low-density liquid and a high-density liquid, where
both liquids are metastable with respect to the diamond cubic (dc) crystal.
Crystallization to the dc crystal always occurs from the low-density liquid.
When $\lambda > 21.5$ crystallization happens so fast that it is no longer
possible to accurately determine the temperature $T_{LL}$ at which the phase
transition occurs for $P=0$.

Simulations of the Stillinger-Weber model indicate that the LLCP for
$\lambda=21$ is located at $-0.60$~GPa and 1120~K \cite{VasishtNP2011}.  Since
each value of $\lambda$ defines a unique system with a unique critical
pressure, the vanishing of the liquid-liquid transition at $\lambda<20.25$
implies that this is the $\lambda$ value for which the LLCP is at $P=0$.
Isochore-crossing studies conducted elsewhere \cite{KapkoPRIVATE2013} show that
this is indeed the case, with $T_c \approx 700$~K for $P_c=0$.  It is clear
that decreasing $\lambda$ means decreasing the tetrahedrality and increasing
density.  When $\lambda<20.25$ the LLCP shifts to positive pressures, and
therefore the phase transition line can no longer be seen in
Fig.~\ref{FIG:modSW}, as it only considers $P=0$.  We thus lack the information
to determine exactly for which $\lambda$ there is no LLCP at {\em any\/}
pressure, but it is certain that this happens at some value $\lambda>0$, since
in the most extreme case of $\lambda=0$ we are left with a simple
Lennard-Jones-like model that has no LLCP.

That weakening the tetrahedrality (i.e., making the tetrahedral bonds more
flexible) leads to the removal of a LLCP, was also shown in 2012 by Tu and
co-authors using a different monatomic model \cite{TuEPL2012}.  The Hamiltonian
of this model includes a term that lowers the energy when particles are aligned
along near-tetrahedral angles and thus favors a diamond cubic ground state.
The study of Ref.~\cite{TuEPL2012} considers two versions: one that allows
broad flexibility of the inter-tetrahedral bond angles (leading to weak
tetrahedrality), and another in which the bond angle is more constrained
(giving rise to strong tetrahedrality).

\begin{figure}[!ht] 
\centering
\includegraphics[angle=0,width=0.5\linewidth]{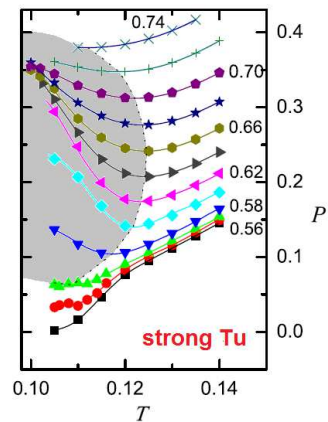}
\caption{ Isochores of the Tu model for the strong tetrahedrality
  version, which has a LLCP
  \cite{TuEPL2012}.
  Gray area indicates the density anomaly
  region.  By reducing the tetrahedrality, the Tu model can be
  smoothly changed into the weak tetrahedrality version, which does
  not have a LLCP.  The isochores of WAC
  (Fig.~\ref{FIG:isochores_BKS_WAC}b) show no LLCP but closely
  resembles that of the strong Tu model.  We can interpret this as
  that WAC is {\em close\/} to having a LLCP, but not close enough.
  If we were to enhance the tetrahedrality of WAC, it is likely a LLCP
  would appear. }
\label{FIG:isochores_strong_Tu}
\end{figure}

The behavior for strong tetrahedrality is shown in
Fig.~\ref{FIG:isochores_strong_Tu}, and we see that the isochores converge into
a critical point.  If the tetrahedrality is weakened slightly, then the
isochores separate, the LLCP disappears, and the diagram starts to resemble
that of Fig.~\ref{FIG:isochores_BKS_WAC}b for WAC.  It should be mentioned that
a separation of the global $C_P$ and $K_T$ maxima also occurs in the weak
tetrahedrality version (as is the case for WAC), while the $C_P$ and $K_T$
maxima are close together and near the LLCP in the strong version of the model.

Finally we should consider the simulations done on ``patchy'' colloids by
Sciortino and coworkers.  Using the Kern-Frenkel (KF) model \cite{KernJCP2003}
(which consists of particles with tetrahedrally arranged sticky points), these
authors demonstrated that the colloids developed tetrahedral network
topologies, with each particle being surrounded by four others---which is not
itself surprising.  More interesting was the finding that, when the effective
sizes of the patches were varied, conditions could be found in which not only
were the relaxation kinetics strictly Arrhenius in form, but also the amorphous
state became the free energy ground state of the system, over a wide range of
densities \cite{SmallenburgNatP2013}.  This corresponds to a more dramatic
stabilization of the amorphous state than the kinetic stability observed in our
work.  It signifies an absolute stability against crystallization on any time
scale, i.e., the system has become an ``ideal glassformer''
\cite{KapkoJCP2013}.

Studies with the KF model have also demonstrated that highly directional bonds
are needed to observe spontaneous crystallization in tetrahedral interacting
particles \cite{RomanoJCP2011}, in agreement with the results found by Molinero
{\it et al.} using the Stillinger-Weber family of potentials.  Since the KF
colloids can be used to describe different tetrahedral models, they promote our
understanding of tetrahedral liquids such as ST2 and mW water, Stilling-Weber
silicon, and BKS silica.  Surprisingly, there exists a mapping from these
models to the KF model, using only a single parameter: the patch width
\cite{SaikaVoivodJCP2013}.  The patch width is related to the flexibility of
the bonds between the particles, and it is therefore likely that spontaneous
crystallization and the existence of a LLCP are related to bond angle
flexibility.

All of these studies show that the occurrence of a LLCP becomes less likely
when the parameters controlling tetrahedrality are weakened.  Unfortunately,
the BKS and WAC models do not have an explicit parameter that controls
tetrahedrality, such as the parameter $\lambda$ in the Stillinger-Weber model.
In this model there is a direct relation between the value of $\lambda$ and the
tetrahedrality of the liquid measured by the orientational order parameter $q$
as defined by Errington and Debenedetti \cite{ErringtonNat2001}.  This
parameter is constructed such that its average value $\left< q \right>$ will
equal zero if all atoms are randomly distributed within the liquid, while $q=1$
for each atom within a perfect tetrahedral network (such as in a cubic diamond
lattice).  For silica the situation is more complicated.  It is not immediately
clear how to define the tetrahedrality of a system that consists of two types
of atoms.  One way would be to find for each Si atom its four nearest
neighboring Si atoms and compute $\left< q \right>$ for this subset of atoms.
However, this measure would completely ignore the positions of the O atoms
which form ionic bridges between the Si atoms.  Since the O-Si-O bond angle
deviates very little from the perfect tetrahedral angle of $109^{\circ}$
\cite{VollmayrPRB1996}, it makes sense to focus on the inter-tetrahedral
Si-O-Si bond angle instead.  It is commonly agreed that structures such as
diamond cubic have maximum tetrahedrality, and for silica this corresponds to a
system where all Si-O-Si bond angles are equal to $180^{\circ}$ (such as
$\beta$-cristobalite).  How much the inter-tetrahedral Si-O-Si bond angles
differ from $180^{\circ}$ can thus be employed as a measure of the
tetrahedrality, and we have therefore calculated this bond angle distribution
for both BKS and WAC.  The location of the maximum in the Si-O-Si bond angle
distribution (i.e., the most probable angle) is a parameter that one could use
to quantify the tetrahedrality.  If we denote the most probable angle at the
lowest accessible temperature ($T_g$) as $\theta_{\text{max}}$, then the
tetrahedrality parameter $t$ can be defined as $t \equiv
\theta_{\text{max}}/180^{\circ}$, where $0 < t < 1$.  Since the ``openness'' of
the structure will increase with the average Si-O-Si angle, one could also
define the tetrahedrality using the volume ratio, i.e., $t \equiv V^{*} /
V_{\text{dc}}$, which would require much less effort to calculate.  Here
$V_{\text{dc}}$ is the volume of the perfect diamond cubic and $V^{*}$ is the
system volume at some corresponding state, for instance at the TMD (which is
less arbitrary than $T_g$).

Let us consider the angular relations and the mechanical forces that determine
them in more detail.  In terms of the familiar ball-and-stick model, the
Si-O-Si bond could be represented by two sticks connected at the oxygen atom,
with a spring in between the sticks.  This spring constrains the bond angle to
some preferred bond angle $\theta_0$, while the value of its spring constant
$k_2$ (the {\em stiffness}) dictates how flexible the bond angle is.  From the
bond angle probability distribution $\mathcal{P}(\theta)$, it is possible to
estimate the values of the preferred bond angle $\theta_0$ and the bond angle
stiffness $k_2$.

\begin{figure}[!t]
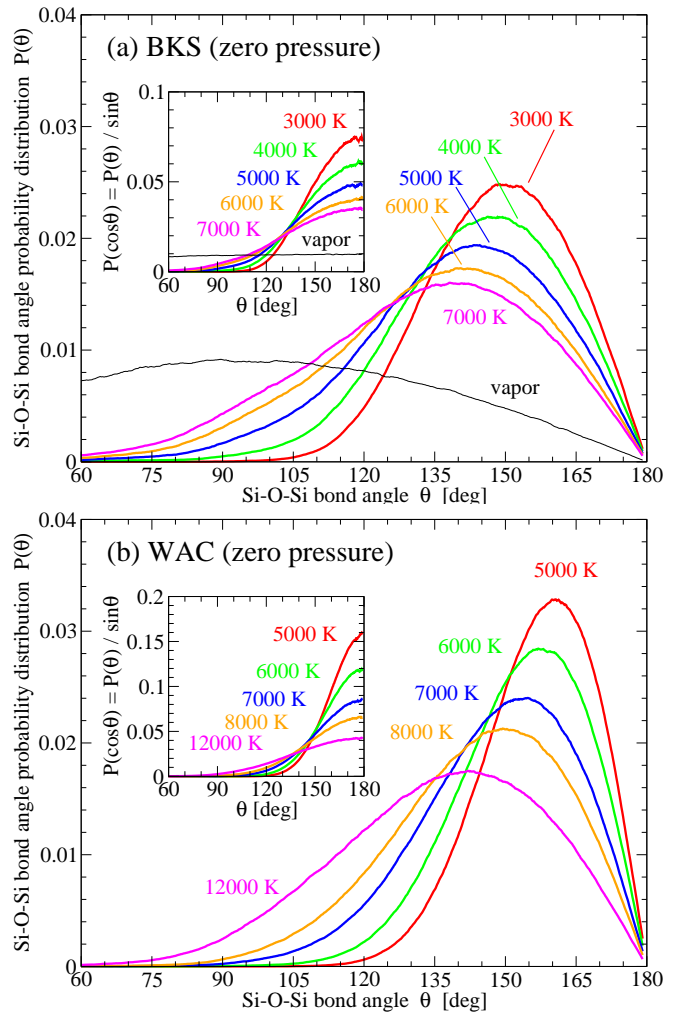
 
\centering
\includegraphics[angle=0,width=\linewidth]{angle-histo-BKS-P0.eps} \\
\includegraphics[angle=0,width=\linewidth]{angle-histo-WAC-P0.eps}
\caption{ Probability distribution of the Si-O-Si bond angle
  $\mathcal{P}(\theta)$ in liquid silica for (a) the BKS model and (b)
  the WAC model.  As $T$ goes down, the most probable angle moves
  closer to $180^{\circ}$ while simultaneously the width of the
  distribution decreases.  The first phenomenon causes the liquid to
  expand upon cooling, while a reduction in width means that the bonds
  become stiffer, which leads to a decrease in diffusion.  Both
  phenomena are related (see below) and are much stronger for WAC than
  for BKS.  Instead of $\mathcal{P}(\theta)$ it is better to consider
  $\mathcal{P}(\cos\theta) = \mathcal{P}(\theta) / \sin\theta$, since
  a completely random distribution such as in the vapor has
  $\mathcal{P}(\theta) \propto \sin\theta$ while
  $\mathcal{P}(\cos\theta)$ is uniform (see inset of panel a).  For
  both models and all temperatures $\mathcal{P}(\cos\theta)$ resembles
  a normal distribution with mean $180^{\circ}$.  This indicates that
  the preferred angle is in fact $180^{\circ}$, and that the width of
  $\mathcal{P}(\cos\theta)$ determines both the location of the peak
  in $\mathcal{P}(\theta)$ as well as its width.  }
\label{FIG:bond_angles}
\end{figure}

To extract the Si-O-Si bond angles from the data, we consider each O ion
together with its two nearest Si neighbors and calculate the angle between the
two Si-O bonds.  In Fig.~\ref{FIG:bond_angles} we show the resulting
probability distributions $\mathcal{P}(\theta)$ of the Si-O-Si angle $\theta$
for BKS and WAC at zero pressure.  These curves have been measured before in
previous studies \cite{VollmayrPRB1996, HemmatiMineralogy2000} but with less
detail.  As the temperature decreases, the width of the distribution decreases
and the maximum shifts toward $180^{\circ}$.  This implies that the liquid
becomes more structured and stiffer.  This is to be expected, since at a high
temperatures there are more thermal fluctuations and therefore
$\mathcal{P}(\theta)$ has a broader distribution.

Plotting $\mathcal{P}(\theta)$ may not be the best way of presenting
the bond angle distribution, as this distribution is biased toward
$90^{\circ}$ angles.  This is particularly clear from the distribution
of the vapor (the thin black line in Fig.~\ref{FIG:bond_angles}a).
The ions in the vapor have no preferred position with respect to their
neighbors, yet $\mathcal{P}(\theta)$ is not uniform but proportional
to $\sin\theta$.  This is related to the fact that the infinitesimal
area element of the unit sphere is $dA = \sin\theta\,d\theta\,d\phi$
rather than $d\theta\,d\phi$.  As $\theta \to 180^{\circ}$ the area
element $dA$ approaches zero, and therefore $\mathcal{P}(\theta) = 0$
at $\theta = 180^{\circ}$.  Instead of $\mathcal{P}(\theta)$ it is
better to consider the probability distribution
$\mathcal{P}(\cos\theta) = \mathcal{P}(\theta)/\sin\theta$, as is
shown in the insets of Fig.~\ref{FIG:bond_angles}.  The
$\mathcal{P}(\cos\theta)$ distribution of the vapor is a uniform
distribution (inset of Fig.~\ref{FIG:bond_angles}a).  For the liquid,
the distribution $\mathcal{P}(\cos\theta)$ is approximately a normal
distribution with its mean at $\theta_0 = 180^{\circ}$.  Evidently the
most probable inter-tetrahedral angle (the location of the
$\mathcal{P}(\theta)$-peak) is purely an effect of the width of this
normal distribution combined with the fact that $dA \propto
\sin\theta$.

It is possible to interpret the bond angle distribution in terms of an
effective potential $U_{\text{eff}}(\theta)$, assuming that
$\mathcal{P}(\cos\theta) \propto \exp[ -U_{\text{eff}}(\theta) /
  k_{B}T ]$.  When the effective potential is harmonic,
i.e. $U_{\text{eff}} = \tfrac{1}{2} k_2 (\theta - \theta_0)^2$, the
resulting probability distribution is a normal distribution with mean
$\theta_0$ and a width that depends on temperature $T$ and stiffness
$k_2$.  In general the effective potential will not be perfectly
harmonic and includes anharmonic terms.  Because $\cos\theta$ is an
even function about $\theta=180^{\circ}$, it is required that
$\mathcal{P}(\cos\theta)$ is as well, and therefore also
$U_{\text{eff}}(\theta)$.  Consequently, the leading-order anharmonic
term in $U_{\text{eff}}(\theta)$ is of the fourth order.  The Si-O-Si
bond angle distribution can thus be described by
\begin{align}
  \mathcal{P}(\theta) = A \sin\theta \exp[ -U_{\text{eff}}(\theta) /
    k_{B}T ]
\label{EQ:Ptheta}
\end{align}
with $U_{\text{eff}}$ a Taylor series about the mean angle $\theta_0 =
180^{\circ}$,
\begin{align}
  U_{\text{eff}}(\theta) = \frac{1}{2} k_2 (\theta-\theta_0)^2 +
  \frac{1}{4!} k_4 (\theta-\theta_0)^4 + \dots
\label{EQ:Ueff}
\end{align}
Here $A$ is a temperature-dependent normalization constant that
ensures that the total probability $\int\mathcal{P}(\theta)\,d\theta =
\int\mathcal{P}(\cos\theta)\,d\!\cos\theta$ is equal to one, and $k_B$
is the Boltzmann constant.

\begin{figure}[!t] 
\centering
\includegraphics[angle=0,width=\linewidth]{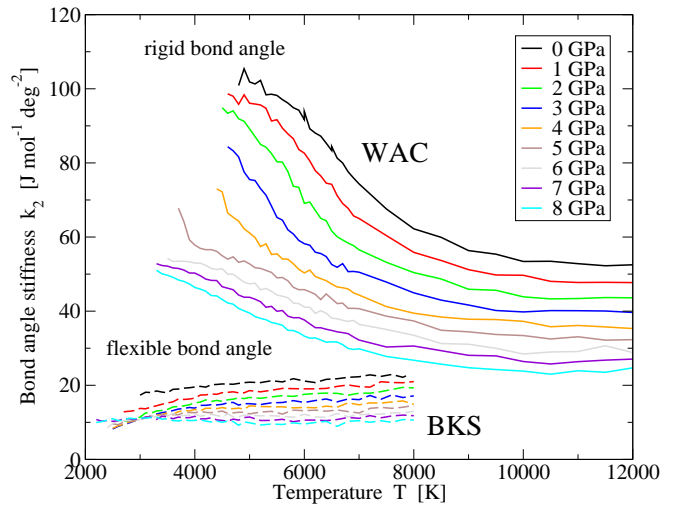}
\caption{ Stiffness of the Si-O-Si bond angle for both WAC (solid
  lines, top) and BKS (dashed lines, bottom).  For both models the
  stiffness $k_2$ goes down with increasing pressure.  It is clear
  that BKS has more flexible bonds (small $k_2$), and that WAC is more
  rigid (large $k_2$) and therefore ``more tetrahedral''.  In addition
  WAC shows a transition at low $T$ for $P \leq 5$~GPa to a state with
  an even higher stiffness.  }
\label{FIG:k2}
\end{figure}

The probability distributions of Fig.~\ref{FIG:bond_angles} can be
fitted quite well with Eqs.~\ref{EQ:Ptheta} and \ref{EQ:Ueff}, even
when the sixth power and higher-order terms are ignored.  The
resulting values for the stiffness $k_2$ are shown in
Fig.~\ref{FIG:k2}.  It is immediately clear that WAC is far more rigid
than BKS.  For BKS the stiffness does not vary much with
temperature, while increasing the pressure makes the bonds slightly
less stiff.
The same is true for WAC at high $T$, but below 5~GPa the
stiffness shows an increase when the liquid is cooled.  This increase
is exactly where $C_P$ has its maximum in
Fig.~\ref{FIG:response_functions_WAC}b, and thus we may argue that the
increase in $C_P$ is due to a structural change, namely the stiffening
of the tetrahedral network.

From the isochores in Fig.~\ref{FIG:isochores_BKS_WAC}b it is clear that WAC is
very close to having a LLCP.  If we compare the results of previous studies
done on tetrahedral liquids \cite{MolineroPRL2006,TuEPL2012} with our results
for BKS and WAC, then we see that the tetrahedrality of BKS is far too small
(i.e., the inter-tetrahedral bond angles are not sufficiently stiff) to have a
LLCP, and that WAC is close, but not close enough.  However, it might be
possible to make a small change to the WAC potential to enhance its
tetrahedrality.  One simple way to achieve this would be to add a repulsive
term similar to the three-body interaction of the Stillinger-Weber model.  This
term should penalize any Si-O-Si configuration with an angle less than
$180^{\circ}$ with a repulsive energy determined by the intensity parameter
$\lambda$ and the size of the deviation.  The $\lambda$ value associated with
this interaction should be carefully chosen; if $\lambda$ is too small no LLCP
will arise, while applying a $\lambda$ that is too large will likely lead to
crystallization into a diamond ($\beta$-cristobalite) structure.  It would be
interesting to see at what value of $k_2$ this criticality is introduced, and
if this value is the same across other tetrahedral models as well, but this is
beyond the scope of the present project.

The results presented here are also relevant to the possible existence of a
LLCP in different water models, and highlight the importance of a thorough
analysis of the O-H-O bond angle distribution.  Such an analysis, possibly with
the use of a bond angle stiffness parameter such as $k_2$, might be able to
predict if a particular water model will have a LLCP.  Unfortunately, to the
best of our knowledge, it is currently not possible to measure these angles
directly in experiments, as significant help from computer simulations is
required to obtain the angular structure of liquid water \cite{SharpACR2010,
SoperPRL2000}.

\section{Conclusion}

\noindent
Although it has been suggested, based on a combination of simulation and
theoretical considerations \cite{SaikaVoivodPRE2000}, that both BKS and WAC
have LLCPs at temperatures beyond the accessible simulation range, our study
suggests that neither BKS nor WAC can reach a critical point.  We have compared
our results to those of other tetrahedral models
\cite{MolineroPRL2006,TuEPL2012}, analyzed the bond angle distributions, and
conclude that the lack of a LLCP in both BKS and WAC is due to a lack of
stiffness in the inter-tetrahedral Si-O-Si bond angles.  WAC is close to
criticality, but BKS shows little sign of a LLCP, and since the latter is
considered to be the more realistic model for experimental silica, we expect
that no LLCP occurs in real silica either.

However, this does not mean that manifestations of criticality can never be
observed.  As Chatterjee and Debenedetti \cite{ChatterjeeJCP2006} have shown
theoretically, even a weak tendency toward criticality (as in BKS) can be
amplified into a liquid-liquid phase separation in a binary system.  Indeed
this notion has been exploited elsewhere \cite{AngellVarna1996} to interpret
the (much-studied \cite{CharlesJACS1966, CharlesJACS1967, CharlesPCG1969,
GalakhovCHAPTER1973, DoremusBOOK1973, HallerJACS1974, MorishitaJACS2004} but
incompletely understood) splitting out of an almost pure SiO$_2$ phase from
such simple systems as the Na$_2$O-SiO$_2$ and Li$_2$O-SiO$_2$ binary glasses
during supercooling.

\section{Acknowledgments}

\noindent
We would like to thank P. Debenedetti, V. Molinero, H. Arag{\~a}o, and C.
Calero for the many valuable discussions.  EL and HES thank the National
Science Foundation (NSF) Chemistry Division for support (Grant No. CHE
12-13217) SVB thanks the Dr. Bernard W. Gamson Computational Science Center at
Yeshiva College for support.  CAA acknowledges the support of this research
through the the National Science Foundation (NSF) experimental chemistry
program under collaborative Grant no. CHE 12-13265.

\appendix
\section{WAC and BKS silica}
\label{appendix_WAC_BKS}

\noindent
One of the simplest models for silica is the WAC model introduced by
L. V. Woodcock, C. A. Angell, and P. Cheeseman \cite{WoodcockJCP1976}.
The model is sometimes also known as the Transferable Ion Model (TRIM)
because its potential is rather general and can also be used to model
other ionic liquids \cite{HemmatiJNCS1997}.  In the WAC model, the
material consists of a 1:2 mixture of Si$^{+4}$ and O$^{-2}$ ions,
without any explicit bonds.  Apart from the electrostatic force, the
ions also interact with each other via an exponential term:
\begin{align}
  & U_{\text{WAC}}(r_{ij}) \equiv \frac{1}{4 \pi \varepsilon_0}
  \frac{z_i z_j e^2}{r_{ij}} + a_{ij} \left(
  1+\frac{z_i}{n_i}+\frac{z_j}{n_j} \right) \times \nonumber \\ &
  \qquad \exp \left[ B_{ij} (\sigma_i + \sigma_j - r_{ij}) \right]
\label{EQ:def_WAC_1}
\end{align}
Here the subscripts $i,j \in \text{Si,O}$ indicate the species of the
two ions involved, $z_i$ the charge of each ion ($z_{\text{Si}}=+4$,
$z_{\text{O}}=-2$), $n_{\text{Si}}=n_{\text{O}}=8$ the number of outer
shell electrons, and $\sigma_i$ the size of each ion
($\sigma_{\text{Si}}=0.1310$~nm, $\sigma_{\text{O}}=0.1420$~nm).  For
WAC silica the parameters $a_{ij}$ and $B_{ij}$ are the same for all
pairs: $a_{ij} = 0.19$~perg $\approx 11.44$~kJ/mol and
$B_{ij}=34.48$~nm$^{-1}$ \cite{HemmatiJNCS1997}.  The potential can
also be written as
\begin{align}
  U_{\text{WAC}}(r_{ij}) = \frac{1}{4 \pi \varepsilon_0} \frac{q_i
    q_j}{r_{ij}} + A_{ij} \exp(-B_{ij} r_{ij}),
\label{EQ:def_WAC_2}
\end{align}
with $A_{\text{SiSi}} =1.917\,991\,469 \times 10^5$~kJ/mol,
$A_{\text{SiO}} = 1.751\,644\,217 \times 10^5$~kJ/mol, and
$A_{\text{OO}} = 1.023\,823\,519 \times 10^5$~kJ/mol.

The second model that we consider here is BKS.  Currently one of the
most popular models, the BKS model was introduced by B. W. H. van
Beest, G. J. Kramer, and R. A. van Santen \cite{vanBeestPRL1990} and
is similar to WAC. Silica is again modeled as a simple 1:2 mixture
of Si- and O-ions, without explicit bonds.  To produce results that
better match experiments and {\it ab initio} simulations, and to be
able to effectively represent screening effects, the charges in BKS
are not integer values of $e$ but instead are given by
$q_{\text{Si}}=+2.4e$ and $q_{\text{O}}=-1.2e$. In addition to this,
the BKS potential also differs from the WAC model in that it includes
an attractive $r^{-6}$ term:
\begin{align}
  U_{\text{BKS}}(r_{ij}) \equiv \frac{1}{4 \pi \varepsilon_0}
  \frac{q_i q_j}{r_{ij}} + A_{ij} \exp(-B_{ij} r_{ij}) - C_{ij}
  r_{ij}^{-6}.
\label{EQ:def_BKS_1}
\end{align}
In BKS there is no interaction between two Si-ions apart from the
electrostatics, i.e. $A_{\text{SiSi}} = B_{\text{SiSi}} =
C_{\text{SiSi}} = 0$.  The parameters for the Si-O pair are
$A_{\text{SiO}} \equiv 18\,003.7572$~eV, $B_{\text{SiO}} \equiv
4.87318$~\AA$^{-1}$, and $C_{\text{SiO}} \equiv 133.5381
$~eV\,\AA$^6$.  For the O-O interaction, the numbers are
$A_{\text{OO}} \equiv 1388.7730$~eV, $B_{\text{OO}} \equiv
2.76$~\AA$^{-1}$, and $C_{\text{OO}} \equiv 175$~eV\,\AA$^6$.

Although the BKS model has been quite successful in simulations of
quartz and amorphous silica, at temperatures above $\sim 5000$K two
ions can come very close, causing problems.
As $r \to \infty$ the BKS potential diverges to $-\infty$ and the two
ions fuse together---a non-physical phenomenon that is an artifact of
the model.  One way to solve this issue is by including an additional
repulsive term at very small $r$, e.g., by adding a $r^{-30}$ term
\cite{SaikaVoivodPRE2000}.  When such a large power is used, however,
a small time step is required to prevent large forces, which leads to
much slower simulations.  Because of this, we instead adjust the BKS
potential at small $r$ by adding a second-degree polynomial for $r\leq
r_{\text{s}}$.  Here $r_{\text{s}}$ is the point at which the original
BKS force has an inflection, i.e., where $d^{2}F_{\text{BKS}}/dr^{2} =
-d^{3}U_{\text{BKS}}/dr^{3} = 0$. We choose the coefficients of the
polynomial such that the new potential $U(r)$ has no inflection at $r
= r_{\text{s}}$. Adding the polynomial still leads to $U(r) \to
-\infty$ when $r \to 0$, but increases the height of the energy
barrier sufficiently to allow us to simulate the high temperatures we
wish to explore.  Choosing a short-range correction to BKS has been
found to have little effect on the simulation results, and merely
prevents the ions from fusing.

To further speed up the simulations, we modify the BKS potential as
described by K. Vollmayr, W. Kob, and K. Binder in
Ref.~\cite{VollmayrPRB1996}, and truncate and shift the potential at
$r_{\text{c}}=0.55$~nm. Although this truncation leads to a shift in
pressure, it otherwise produces approximately the same results
\cite{VollmayrPRB1996}.  In conclusion, the modified BKS potential we
use is given by
\begin{align}
  & U'_{\text{BKS}}(r_{ij}) = \frac{1}{4 \pi \varepsilon_0}\frac{q_i
    q_j}{r_{ij}} \nonumber \\ &+ \left\{
    \begin{array}{lll}
    a_{ij} r_{ij}^2 + b_{ij} r_{ij} + c_{ij} - \frac{1}{4 \pi
      \varepsilon_0}\frac{q_i q_j}{r_{ij}} & & (r_{ij} < r_{\text{s}})
    \\ A_{ij} \exp(-B_{ij} r_{ij}) - C_{ij} r_{ij}^{-6} -
    U_{\text{c},ij} & & (r_{\text{s}} < r_{ij} < r_{\text{c}}) \\ 0 &
    & (r_{ij} > r_{\text{c}}), \\
    \end{array}
  \right.
\label{EQ:def_mod_BKS}
\end{align}
with the parameter values for $ij=\text{SiO}$ and $ij=\text{OO}$
listed in Table~\ref{TAB:parameters_mod_BKS}.  For the Si-Si
interaction the potential is $U'_{\text{BKS}}(r_{\text{SiSi}}) =
\frac{1}{4 \pi \varepsilon_0} q_{\text{Si}}^2 /r_{ij}$ and does not
involve any cutoffs, apart from the real-space cutoff of the Ewald
sum.

\begin{table}[t] 
\centering
\begin{tabular}{l|r|r|l}
                     &  Si-O                               &  O-O                                &  units  \\
\hline
  $a_{ij}$           &   2.678\,430\,850$\times$10$^{5}$   &   9.208\,901\,230$\times$10$^{4}$   &  kJ/mol\,nm$^2$  \\
  $b_{ij}$           & $-7.343$\,377\,221$\times$10$^{4}$  & $-4.873$\,373\,066$\times$10$^{4}$  &  kJ/mol\,nm  \\
  $c_{ij}$           &   2.353\,960\,789$\times$10$^{3}$   &   7.337\,042\,047$\times$10$^{3}$   &  kJ/mol  \\
  $A_{ij}$           &   1.737\,098\,076$\times$10$^{6}$   &   1.339\,961\,920$\times$10$^{5}$   &  kJ/mol  \\
  $B_{ij}$           &  48.7318                            &  27.6                               &  nm$^{-1}$  \\
  $C_{ij}$           &   1.288\,446\,484$\times$10$^{-2}$  &   1.688\,492\,907$\times$10$^{-2}$  &  nm$^6$\,kJ/mol  \\
  $U_{\text{c},ij}$  & $-0.465\,464\,470$                  & $-0.575\,753\,031$                  &  kJ/mol  \\
  $r_{\text{s}}$     &   0.139\,018\,528                   &   0.195\,499\,453                   &  nm  \\
  $r_{\text{c}}$     &   0.55                              &   0.55                              &  nm  \\
\end{tabular}
\caption{ Parameters of the modified BKS potential of
  Eq.~(\ref{EQ:def_mod_BKS}).  Because Si-Si only has the (repulsive)
  Coulomb interaction, all parameters are zero for Si-Si.  One mol
  here indicates one mol of ions, not one mol of SiO$_2$ molecules.  }
\label{TAB:parameters_mod_BKS}
\end{table}

\section{Calculation of response functions via surface fits}
\label{appendix_smooth_surface}

\noindent
In order to construct isobaric response functions from a large set of
constant-volume ($NVT$) data, some type of fit or interpolation is
needed.  For example, to calculate $C_P = (\partial H / \partial T)_P$
we consider the enthalpy $H$ as a function of both $P$ and $T$ and fit
the data $[P,T,H]$ with a smooth 3-dimensional surface $H(P,T)$.
Abrupt changes in $H(P,T)$ lead to large spikes in its derivative
$\partial H / \partial T$, and thus the $H(P,T)$ surface must be
smooth if we are to obtain a meaningful $C_P$.  Fitting a surface
rather than a curve has the additional advantage that more data is
taken into account, resulting in better statistics.  An alternative
approach is to calculate $C_P$ via fluctuations in $H$, but it has
been shown \cite{LascarisAIP2013} that first fitting $H(T)$ and then
taking a derivative leads to cleaner results.  It is of course easier
to calculate $C_P$ by doing constant-pressure ($N\!PT$) simulations
instead, but then one would have the same problem with calculating
$C_V$.  We conclude that we can easily calculate all response
functions if we apply a smooth surface fit $f(x,y)$ to a set of
3-dimensional points $z_k(x_k,y_k)$.

Fitting a surface to a set of points means striking a balance between
the ``smoothness'' of the fit and the fitting error induced.  One
measure of smoothness is the Laplacian $\nabla^2 f$, since a small
Laplacian means little change in the slope of $f(x,y)$, and thus a
smoother function.  Hence, to obtain a smooth surface fit $f(x,y)$
through the data points $z_k(x_k,y_k)$ with $k = 1,2,\dots,N$, we
minimize
\begin{align}
  J = \sum_{k=1}^{N} w_k \left[ f(x_k,y_k) - z_k \right]^2 + \iint
  \left| \nabla^2 f(x,y) \right|^2 dx\,dy.
\label{EQ:smooth_surface}
\end{align}
The weights $w_k$ provide the balance between the smoothness and the
fitting error.  If we set $w_k$ too low, we obtain a very smooth
fitting function $f(x,y)$ that poorly represents the data. If we set
$w_k$ too high, the function $f(x,y)$ will go through all the data
points but will show large variations.  Because large variations in
the surface lead to even larger variations in the derivatives, the
$H(P,T)$ surface must be very smooth when we calculate the $C_P$.
Fortunately, introducing small fitting errors does not cause problems,
because the simulation data already suffers from small statistical
errors. If the underlying response function is in fact smooth, then it
is possible to use the fitting errors to partially cancel the
statistical errors.

Minimization of the functional $J$ in Eq.~\ref{EQ:smooth_surface} is
not a new concept.  For example, the \textsc{csaps} function in MATLAB
applies a similar minimization scheme to calculate a cubic smoothing
spline.  As opposed to this MATLAB function, we do not impose the
constraint that $f(x,y)$ is a tensor product spline, but instead
represent $f(x,y)$ by a set of $100 \times 100$ points
$(x_i,y_j,f_{ij})$ placed on a regular grid $(x_i,y_j)$.  Bilinear
interpolation is used to estimate the value of $f(x,y)$ between these
grid points, and the derivatives and the Laplacian are calculated
using finite (central) differences.  To compensate for the reduced
number of data points near the edges of the domain, we recommend that
higher-order differences near the edges be used.


\bibliographystyle{jcp}  


\end{document}